\documentstyle[12pt]{article}
\begin{document}
\newcommand{\bea}{\begin{eqnarray}}
\newcommand{\eea}{\end{eqnarray}}
\newcommand{\ba}{\begin{array}}
\newcommand{\ea}{\end{array}}
\newcommand{\eee}{\mbox{e}}
\newcommand{\triex}{\mbox{\hspace{3ex}}}
\newcommand{\ovl}{\overline}
\newcommand{\edoc}{\end{document}}
\newcommand{\pal}{\partial}
\newcommand{\sta}{\stackrel}
\newcommand{\ch}{\mbox{ch\,}}
\newcommand{\sh}{\mbox{sh\,}}
\newcommand{\th}{\mbox{th\,}}
\newcommand{\spsigma}{\ba[t]{c} \mbox{Sp} \vspace{-1ex} \cr
\mbox{$\scriptstyle{(\sigma)}$} \ea }

%
%
%
%
%

\title{\LARGE Fermionic Structure of Two-Dimensional Ising Model
with Quenched Site Dilution}

\author{
V.N. Plechko \\[1.5ex]
\em  Bogoliubov Laboratory of Theoretical Physics, \\
\em  Joint Institute for Nuclear Research, JINR-Dubna, \\
\em  141980 Dubna, Moscow Region, Russia
\thanks{ \ E-mail address: plechko@thsun1.jinr.dubna.su} }

\maketitle

%
%
\begin{abstract}
\noindent We apply a new anticommuting path integral technique to clarify
the fermionic structure of the 2D Ising model with quenched site dilution.
In the $N$-replica scheme, the model is explicitly reformulated as a theory
of interacting fermions on a lattice. An unusual feature is that the
leading term of interaction in the exact lattice theory is of order $2N$ in
fermions, where $N$ is the number of replicas. The continuum-limit
approximation near $T_c\,$ for weak dilution produces, however, an
effective four-fermion interaction. In particular, this implies the
doubled-logarithmic singularity in the specific heat near $T_c$ for weak
site dilution. The exact value of the initial slope is also obtained.  \end{abstract}
\thispagestyle{empty}

\newpage
\setcounter{page}{1}

%
%
\section{Introduction}

{}From the experimental point of view, the site dilution provides,
probably, the most simple way to realize quenched disorder in real
magnetic materials. In this case, some amount of the magnetic atoms in a
sample, chosen at random, are replaced by nonmagnetic impurity atoms.
Another sort of disorder is bond dilution, merely preferred in theoretical
studies, in which case some of the lattice bonds are assumed to be broken.
The two-dimensional Ising model (2DIM) is a natural object to analyze the
effects of disorder since it admits the exact analytic solution in the pure
case \cite{ons44}.   This model, on the other hand, is too delicate to
make definite predictions on  common phenomenological grounds. In
particular, from the point of view of the Harris criterion
\cite{har74,stinch83} (which states that the critical behaviour is modified
by quenched disorder if $\alpha>0$, and is unchanged if $\alpha<0$, where
$\alpha$ is the specific heat exponent for the pure system) the 2D Ising
model is marginal, since $\alpha=0$ in the pure case. The 2D Ising model
with quenched dilution has been intensively analyzed during the last
decades both by theoretical methods \cite{har74}-\cite{jugsh96} and in
the precise Monte-Carlo experiments \cite{wsdoa90p}-\cite{SST95}. The
theoretical studies here were merely concerned with bond dilution, making
use of the fermionic interpretation of the Ising model
\cite{dd81}-\cite{jugsh96}. The case of site dilution, however, has not yet
been properly analyzed within the fermionic approach. The site dilution can
be viewed as locally correlated bond dilution. This makes the traditional
transfer-matrix and combinatorial methods of the fermionization, which is
the basis for theoretical analysis, less suitable in the case of site
disorder \cite{jugsh96}. Meanwhile, the Monte-Carlo studies \cite{kim94}
and related analyses \cite{kuhn94} give evidence for a different behaviour,
at least for moderated and strong site dilution, from that predicted in the
theoretical studies of weak bond dilution \cite{dd83,shal94}. The
alternative theories for bond dilution are also to be taken into account
\cite{timo89,zieg90}. A number of questions thus arise about the
universality between site and bond dilution, and between weak and strong
dilution in both the cases \cite{SST95}-\cite{ruiz97}.  Also, see
\cite{SST95} for a critical review.

In this Letter we apply a novel anticommuting path integral technique to
clarify the fermionic structure of the two-dimensional Ising model with
quenched site dilution. At the first stage, we transform the partition
function of the 2D Ising model with fixed site dilution into a Gaussian
fermionic integral. The averaging over the disorder in the $N$-replica
scheme then yields a lattice theory with compact multifermion interaction.
An interesting feature is that the term of interaction here is of order
$2N$ in fermions, where $N$ is the number of replicas. Notice that the
formal limit $N \to0$ is assumed at final stages. For weak dilution,
however, the effective continuum-limit theory responsible for critical
behaviour near $T_c$ appears to be the standard $N=0$ Gross-Neveu model
with four-fermion interaction, already well known in the DD-SSL theories
for weak bond dilution \cite{dd83,shal94}. In particular, this
implies the doubled-logarithmic singularity in the specific heat and
log-corrections in other thermodynamic functions near $T_c$ for weak site
dilution. An effective four-fermion term arises in the continuum theory, in
our case, due to the interplay of the short-wave and long-wave lattice
fermionic modes. The parameters of the $N=0$ Gross-Neveu model for site
dilution are then explicitly evaluated.  These parameters include some
characteristic lattice fermionic averages for the pure Ising model at
$T_c$. The exact value for the initial slope of the $T_c\,(p)$-$p$ curve
is also obtained.  The basic results are given in (\ref{qab1}),
(\ref{qcc2}), (\ref{sint1}), (\ref{gross1}), (\ref{aveab2}). Further
comments are added in the following sections.

%
%
\section{The model}

Let us start with the two-dimensional Ising model with fixed distribution
of the nonmagnetic sites over the lattice (fixed site dilution). At the
first stages, we also assume arbitrary inhomogeneous distribution of the
coupling parameters over the lattice bonds since this will not be an
obstacle for fermionization. The hamiltonian is:
\bea
-\beta H =\mbox{$\sum\limits_{mn}^{}$}
\,[\,b_{m+1n}^{\,(1)}\,y_{mn}y_{m+1n}\sigma_{mn}\sigma_{m+1n}
+ b_{mn+1}^{\,(2)}\,y_{mn}y_{mn+1}\sigma_{mn}\sigma_{mn+1}\,]\,,
\label{ham1}
\eea
where the Ising spins, $\sigma_{mn}=\pm1$, are disposed at the lattice
sites, $mn$, with $m,n=1,2,...,L\,$, $L$ is the lattice length. To
introduce site dilution, we supply each Ising spin by random variable
$y_{mn}=0,1$ playing the role of the effective magnetic moment of
the Ising spin. If $y_{mn}=1$, the given site is normal, if $y_{mn} =0$,
the site is dilute.  Finally,  $b_{mn}^{\,(\alpha)} = \beta J_{mn}^{\,
(\alpha)}$ are the dimensionless bond couplings, $J_{mn}^{\,(\alpha)}$ are
the exchange energies, $\beta=1/kT$ is the inverse temperature in the
energy units.  $J_{mn}^{\,(\alpha)}>0$ corresponds to the ferromagnetic
bond. For fixed disorder, the partition function and free energy are:
\bea
Z\,\{y\}\,=\sum\limits_{\,(\sigma)}^{}\,
\eee^{\,-\beta H\,\{y\,|\,\sigma\}}=
\eee^{\,-\beta F\,\{y\}}\,,
\label{zet1}
\eea
where the sum is taken over all the possible spin configurations provided
by $\sigma _{mn} =\pm1$ at each site. Taking into account the identity for
a typical Boltzmann  weight: $\exp\,(b\,yy'\sigma\sigma') =\cosh(b\,yy')
(1+yy' \sigma\sigma'\,\tanh b\,)$, which readily follows from
$\sigma\sigma' =\pm1$, $yy'=0,1$, the partition function can be written in
the form:  $Z\,\{\,y\,\}=R\, \{\,y\,\}\,Q\,\{\,y\,\}$, where $Q\{y\}$ is
the reduced partition function and $R\{y\}$ is a nonsingular
spin-independent prefactor to be ignored in what follows. The reduced
partition function is:
\bea
Q\,\{y\} =\spsigma\!\prod\limits_{mn}^{}
(\,1+t_{m+1n}^{\,(1)}\, y_{mn}y_{m+1n}\sigma_{mn}\sigma_{m+1n})\,
(1+t_{mn+1}^{\,(2)}\,y_{mn}y_{mn+1}\sigma_{mn}\sigma_{mn+1})\,,
\label{qss1}
\eea
where $t_{mn}^{\,(\alpha)}= \tanh\,b_{mn}^{\,(\alpha)}$ and we assume a
properly normalized spin averaging such that ${\rm Sp}\,(1)=1,\, {\rm
Sp}\, (\sigma_{mn})=0$ at each site.

Since we are interested in quenched disorder, the averaging over the
random-site (RS) impurities is to be performed for the free energy rather
than for the partition function itself. The standard device to avoid the
averaging of the logarithm is the replica trick:
\bea
&&  \ovl{\stackrel{}{\ln\,Q\{\,y\,\}}}
= \lim_{N \to 0}\,\frac{1}{N}\,
\ovl{\stackrel{}{(Q^N\{y\}-1)}}\,.
\label{repl1}
\eea
In this scheme, we take $N$ identical copies of the original
partition function and average $Q^{\,N}\{y\}\,$. The formal limit
$N\to0$ is assumed at final stages. We will also assume the following
distribution of the RS impurities in the definition of the averaging:
\bea
W\,(y_{mn}) = p\,\delta\,(1 - y_{mn}) + (1 - p)\,\delta(y_{mn})\,,
\label{wwy1}
\eea
where $\delta(\ )$ are the correspondent Kronecker's symbols, $p$ is the
probability that any given site, chosen at random, is occupied by the
normal Ising spin, while $1-p$ is the probabilty that the given site is
dilute.

%
%
\section{Fermionization}

Before averaging over the disorder, we have to transform $Q\{y\}$
into a fermionic Gaussian integral. Let us remember that Grassmann
variables (nonquantum fermionic fields) are the purely anticommuting
fermionic symbols. Given a set of Grassmann variables $a_1,a_2,...\,, a_N$,
we have $a_ia_j+ a_ja_i =0$, $\,a_{j}^{2} =0$. The notion of the integral
over Grassmann variables (fermionic path integral) was first introduced by
Beresin \cite{ber66}. The elementary rules of integration for one variable
are \cite{ber66}:
\bea
\int da_j \cdot a_j = 1\,, \triex
\int da_j \cdot 1   = 0\,. \triex
\label{gr2}
\eea
In a multidimensional integral, the differential symbols $da_1,da_2,
\ldots,da_N$ are again anticommuting with each other and with the
variables. The fermionic exponential is assumed in the sense of its series
expansion. For a finite number of the variables, the exponential series
definitely terminates at some stage due to the property $a_{j}^{2}=0$. In
the field-theoretical language, the fermionic form in the exponential is
called action.

The transformation of $Q\{y\}$ into a fermionic integral can be realized
by following literally the fermionization procedure first developed for the
pure $2D$ Ising model \cite{ple85d}-\cite{ple88}. For a recent comment
see also \cite{ple95,ple96}. The first step is the fermionic factorization of
the local bond Boltzmann weights  \cite{ple85d,ple85t}.  For the whole
lattice, we introduce a set of the totally anticommuting Grassmann
variables, $a_{mn}, a_{mn}^{\,*}, b_{mn}, b_{mn}^{\,*}$, and write:
\bea
&& 1+t_{m+1n}^{(1)}y_{mn}^{}y_{m+1n}^{}\sigma_{mn}\sigma_{m+1n}
\cr
&& =\int\limits_{}^{}da_{mn}^{*}da_{mn}\,\mbox{e}^{\,a_{mn}a_{mn}^{*}}\,
(1+a_{mn}^{}y_{mn}^{}\sigma_{mn}^{})\,
(1+t_{m+1n}^{(1)}\,a_{mn}^{*}y_{m+1n}^{}\sigma_{m+1n}^{})\,,
\cr\cr
&& 1+t_{mn+1}^{(2)}y_{mn}^{}y_{mn+1}^{}\sigma_{mn}\sigma_{mn+1}
\cr
&& =\int\limits_{}^{}db_{mn}^{*}db_{mn}\,\mbox{e}^{\,b_{mn}b_{mn}^{*}}\,
(1+b_{mn}^{}y_{mn}^{}\sigma_{mn}^{})\,
(1+t_{mn+1}^{(2)}\,b_{mn}^{*}y_{mn+1}^{}\sigma_{mn+1}^{})\,.
\label{fact1}
\eea
These identities can be readily checked using the elementary rules like
(\ref{gr2}) and taking into account that $\eee^{\,aa^{\,*}} = 1 +
aa^{\,*}$, since $(aa^{\,*}) ^{\,2} =0$. Neglecting the sign of the Gaussian
fermionic averaging, the weights are now presented as $A_{mn}^{}
A_{m+1n}^{\,*}\,$, $B_{mn}^{} B_{mn+1} ^{\,*}$, where the separable factors
(to be called Grassmann factors) are to be identified from (\ref{fact1}).
At the next stage we group together, over the whole lattice, the four
factors with the same Ising spin (the same index $mn$) and then sum over
$\sigma_{mn}=\pm1$ in each group of factors independently, thus passing to
a purely fermionic expression for $Q\{y\}$. The above four factors with the
same $\sigma_{mn}$ come by factorization of the four bonds attached to the
given $mn$ site, they all include also the same variable $y_{mn}$. There is
also the ordering problem, to be solved, for the noncommuting Grassmann
factors in their global products. This problem is related to the
requirement that we can actually keep nearby the four factors with the same
spin when averaging over $\sigma_{mn}=\pm1$ at each site.  In the $2D$
case, the ordering problem is resolved in terms of the `mirror-ordered
factorization procedure' for the global products of weights
\cite{ple85d}-\cite{ple88}.  The appearance (as compared with the pure
case) of additional parameters $y_{mn}^{}$ in the factors arising in
(\ref{fact1}) does not matter in this respect. The averaging over
$\sigma_{mn} \pm1$ in the local group of four factors yields an even
fermionic polynomial. This polynomial can be represented as an exponential
factor associated with the $mn$ lattice site.  This is just the exponential
factor corresponding to the $y_{mn}^{2}$ term in (\ref{qab1}) below. Thus
we come to the result:
\bea
&& Q\,\{y\}\;=\; \int \prod\limits_{mn}^{}db_{mn}^{\,*}db_{mn}^{}
da_{mn}^{\,*}da_{mn}^{}\;\exp\,\sum\limits_{mn}^{}\,
\Big\{\,a_{mn}^{}a_{mn}^{\,*}+b_{mn}^{}b_{mn}^{\,*}
\nonumber \\
&& +\,y_{mn}^{2}\,[\,a_{mn}^{}b_{mn}^{}+\,t_{mn}^{(1)}t_{mn}^{(2)}
a_{m-1n}^{\,*}b_{mn-1}^{\,*}+\,(t_{mn}^{(1)}a_{m-1n}^{\,*}+
t_{mn}^{(2)}b_{mn-1}^{\,*})\,(a_{mn}^{}+b_{mn}^{})\,]\,\Big\}\,.
\label{qab1}
\eea
In the RS pure case ($y_{mn} \equiv 1$) we come back to Eq. (11) of Ref.
\cite{ple85d} (inhomogeneous distribution of bond couplings was already
assumed in \cite{ple85d}). In fact, noting (\ref{fact1}), one can guess
(\ref{qab1}) directly from Eq. (11) of Ref. \cite{ple85d} without special
calculation. The partition function with fixed disorder is represented now
as a Gaussian fermionic integral.

The integral (\ref{qab1}) can be in turn simplified integrating out the
$a_{mn}, b_{mn}$ fields by means of the identity $\int db\,da\,
\exp\,(\lambda\,ab + aL + L^{\,'}b) = \lambda + LL' =\lambda\exp\,
\big(\,\lambda^{\,-1}\,LL'\,\big)$, where $a,b$  are Grassmann variables,
$L,L^{\,'}$ are linear fermionic forms independent of $a,b$, and
$\lambda$ is a parameter \cite{ple96}. For $\lambda =0$ the result is
$LL'$. Integrating out the $a_{mn},b_{mn}$ fields in this way, we obtain a
reduced integral in terms of variables $a_{mn}^{\,*},b_{mn}^{\,*}$.
Changing the notation for the fields, for the sake of visual comfort, as
follows: $a_{mn}^{*}, b_{mn}^{*} \to c_{mn}, -\,\bar{c}_{mn}$, we find the
reduced integral in the form:
\bea
Q\,\{y\}= \int \prod\limits_{mn}^{} d\bar{c}_{mn}^{}
dc_{mn}^{}\,y_{mn}^{\,2}\,\exp\sum\limits_{mn}^{}\,
\Big[\,y_{mn}^{-2}c_{mn}^{}\bar{c}_{mn}^{}-\,y_{mn}^{\,2}\,
t_{mn}^{(1)}t_{mn}^{(2)}\, c_{m-1n}^{}\bar{c}_{mn-1}^{}
\cr
+\,(c_{mn}^{}+\bar{c}_{mn}^{})\,(t_{mn}^{(1)}\,c_{m-1n}^{} -\,
t_{mn}^{(2)}\,\bar{c}_{mn-1}^{})\,\Big]\,,
\label{qcc1}
\eea
where $y_{mn}^{\,2}\exp\,(y_{mn}^{-2}c_{mn}^{}\bar{c}_{mn})= y_{mn}^{\,2}
+c_{mn}\bar{c}_{mn}\,$, there is no true singularity at $y_{mn}^{2}=0$.
The fermionic integrals (\ref{qab1}) and (\ref{qcc1}) are equivalent to
each other and to (\ref{qss1}). This equivalence holds even independently
of the formal interpretation of the variables $y_{mn}$. However, we have
essentially used the property $y_{mn}^{}=0,1$ by passing from $Z\{y\}$ to
$Q\{y\}$. If $y_{mn}^{2}=0,1$, then one can put $y_{mn}^{}$ instead of
$y_{mn}^{2}$ in the above integrals. In what follows we assume definitely
that $y_{mn}=0,1$. The structure of the integral can be illuminated then
most obviously by expanding the local factor in (\ref{qcc1}) over the
`eigenstates' with $y_{mn}=0$ and $y_{mn}=1$:
\bea
Q\,\{y\} = \int \prod\limits_{mn}^{} d\bar{c}_{mn}^{} dc_{mn}^{}\,
\prod\limits_{mn}^{}\,\Big[\,\delta\,(1-y_{mn}^{})\,\exp\,(S_{mn})
+ \delta\,(y_{mn}^{})\, c_{mn}^{}\bar{c}_{mn}^{}\,\Big]\,,
\label{qcc2}
\eea
where $\delta\,(1-y_{mn})=y_{mn}$ and $\delta\,(y_{mn})=1-y_{mn}$ are
the correspondent Kronecker symbols, $y_{mn}=0,1$, and  $S_{mn}$ is
the local density of action for the RS pure case, see (\ref{qpp1}).

In the RS pure case, with $y_{mn}\equiv 1$ at all sites, we obtain:
\bea
Q\,\{1\}=\int \prod\limits_{mn}^{}D_{mn}\,
\exp\,\sum\limits_{mn}^{}S_{mn}
= \int \prod\limits_{mn}^{} d\bar{c}_{mn}dc_{mn}\,\exp\,
\sum\limits_{mn}^{}\,\Big[\,c_{mn}\bar{c}_{mn}
\nonumber \\
+\,(c_{mn} +\bar{c}_{mn})\,(t_{mn}^{(1)}c_{m-1n}
-\,t_{mn}^{(2)} \bar{c}_{mn-1})
-\,t_{mn}^{(1)}t_{mn}^{(2)}c_{m-1n}\bar{c}_{mn-1}\,\Big]\,,
\label{qpp1}
\eea
where $D_{mn}$ and $S_{mn}$ are the abbreviations for the local terms in
the fermionic measure and the action, respectively. The inhomogeneous
distribution of bond couplings is still preserved. Therefore, integral
(\ref{qpp1}) can be used as a starting point to analyze bond dilution. We
can assume as well that random bonds may be antiferromagnetic, thus making
connection to spin glasses. The discussion of these subjects, however, is
out of the scope of the present exposition.

%
%
\section{ The N-replicas (lattice)}

Within the $N$-replica scheme, see (\ref{repl1}), we have to average the
replicated partition function. From (\ref{qcc2}), the replicated partition
function is:
\bea
&& Q_{}^{\,N}\{y\} =
\int\,\prod\limits_{\alpha=1}^{N}\,\prod\limits_{mn}^{}\,
d\bar{c}_{mn}^{\;\alpha}\,dc_{mn}^{\,\alpha}\,
\prod\limits_{mn}^{}
\Big[\,\delta\,(1-y_{mn}) \prod\limits_{\alpha=1}^{N}\eee^{\,
S_{mn}^{\,\alpha}} + \delta\,(y_{mn})\,\prod\limits_{\alpha=1}^{N}
c_{mn}^{\,\alpha}\bar{c}_{mn}^{\;\alpha}\,\Big]\,,\triex
\label{qnn30}
\eea
where $S_{mn}^{\,\alpha}$ ($\alpha=1,\ldots,N$) are the replicas of the
RS-pure local action from (\ref{qpp1}). The averaging over the
disorder with distribution (\ref{wwy1}) then yields:
\bea
&&  \ovl{\stackrel{}{Q^N\{y\}}}=
\int \prod\limits_{\alpha=1}^{N}\prod\limits_{mn}^{}
d\bar{c}_{mn}^{\;\alpha}dc_{mn}^{\,\alpha}\,\prod\limits_{mn}^{}
\Big[\; p\,\prod\limits_{\alpha=1}^{N}
\eee^{\,S_{mn}^{\,\alpha}} + (1-p)\,\prod\limits_{\alpha=1}^{N}
c_{mn}^{\,\alpha}\bar{c}_{mn}^{\;\alpha}\;\Big]\,
\nonumber  \\
&&
=\,p^{L^2}\int\prod\limits_{\alpha=1}^{N}\prod\limits_{mn}\,
d\bar{c}_{mn}^{\,\alpha}dc_{mn}^{\,\alpha}\,\exp\,
\sum\limits_{mn}^{}\Big[\sum\limits_{\alpha=1}^{N}\,
S_{mn}^{\,\alpha} + \frac{1-p}{p}\,\prod\limits_{\alpha=1}^{N}
c_{mn}^{\,\alpha}\bar{c}_{mn}^{\;\alpha}\,
\exp\,(\,-\,S_{mn}^{\,\alpha})\;\Big]\,
\nonumber \\
&&=\,p^{L^2}\int\prod\limits_{\alpha=1}^{N}\prod\limits_{mn}\,
d\bar{c}_{mn}^{\;\alpha}dc_{mn}^{\,\alpha}\,\exp\,
\sum\limits_{mn}^{}\Big[\,\sum\limits_{\alpha=1}^{N}\,
S_{mn}^{\,\alpha}+ \frac{1-p}{p}\,\prod\limits_{\alpha=1}^{N}
c_{mn}^{\,\alpha}\bar{c}_{mn}^{\;\alpha}\,\exp\,
(\Delta_{mn}^{\,\alpha})\;\Big]\,.
\label{qnn50}
\eea
We elaborate the product from the first line into a unique exponential
making use of the nilpotent property of fermions, $(c_{mn}^{\,\alpha})^{2}=
(\bar{c} _{mn}^{\,\alpha})^{2}=0$. In the final line, referring to the same
property, we replace $-S_{mn}^{\alpha}$ by the reduced form,
$\Delta_{mn}^{\alpha} =t_{mn}^{(1)}t_{mn}^{(2)}\,c_{m-1n}^{\,\alpha}
\bar{c}_{mn-1}^{\,\alpha}\,$, taking into account that prefactor
$c_{mn}^{\,\alpha}\bar{c}_{mn} ^{\,\alpha}$ annihilates the correspondent
fermions in $\exp\,(-S_{mn}^{\alpha})$. Yet another possibility, preferably
to be used below, is the choice: $\Delta_{mn}^{\,\alpha}=
t_{mn}^{(1)}t_{mn}^{(2)} (\pal_m c_{mn}^{\,\alpha})\, (\pal_n
\bar{c}_{mn}^{\,\alpha})\,$, where $\pal_m,\pal_n$ are lattice
derivatives:  $\pal_{m}\,c_{mn}^{\,\alpha}= c_{mn}^{\,\alpha}
-c_{m-1n}^{\,\alpha}$, $\pal_{n}\,\bar{c}_{mn}^{\,\alpha} =
\bar{c}_{mn}^{\,\alpha} - \bar{c}_{mn-1} ^{\,\alpha}\,$.

We have thus formulated the exact lattice fermionic theory for the 2D
Ising model with quenched site dilution. The integral resulting in
(\ref{qnn50}) can be written shortly in the form:
\bea
\ovl{\stackrel{}{\mathstrut Q^{N}\{y\}}}
=\int D\,\exp\,(S_0 + \lambda_0\,S_{\rm\,int})\,, \;\;\;\;\;
\lambda_0= \frac{1-p}{p}\,,
\label{qnn80}
\eea
where $D$ is the replicated fermionic measure, $S_0$ is the Gaussian part
of the action, which is the replicated action for the RS pure case, eq.
(\ref{qpp1}), while the non-Gaussian term of interaction can be written, in
particular, as follows:
\bea
\lambda_0\, S_{\rm\,int} =\lambda_0\,\sum\limits_{mn}^{}\Big\{
\prod\limits_{\alpha=1}^{N} c_{mn}^{\,\alpha}\bar{c}_{mn}^{\,\alpha}\,
\exp\,(t_{mn}^{(1)}t_{mn}^{(2)}\,(\pal_m c_{mn}^{\,\alpha})(\pal_n
\bar{c}_{mn}^{\,\alpha})\,\big)\Big\}
\nonumber \\
= \lambda_0\,\sum\limits_{mn}^{}\Big\{ \prod\limits_{\alpha=1}^{N}
c_{mn}^{\,\alpha}\bar{c}_{mn}^{\,\alpha}\,
\Big[\,1+t_{mn}^{(1)}t_{mn}^{(2)}\,(\pal_m c_{mn}^{\,\alpha})(\pal_n
\bar{c}_{mn}^{\,\alpha})\,\Big] \Big\}\,. \;\;\;
\label{sint1}
\eea
The interaction appears to be of order $2N$ in fermions, where $N$ is the
number of replicas, with $N\to 0$ at final stages. This is a somewhat
unusual feature of the resulting model. In a sense, in (\ref{sint1}), at
each lattice site we have the product of `holes' (virtual dilute sites)
over the replicas. Respectively, the coupling constant is of order $1-p$,
which is a fraction of the dilute sites over the lattice.  Despite some
unexpected features, the interaction (\ref{sint1}) is of a relatively
simple and compact form. The resulting fermionic theory can be used,
probably, in future research, for the analysis of strong and moderated
dilution (it might be important to take properly into account short lattice
distances).  For weak dilution near $T_c$, where the continuum-limit
approximation can be applied, the standard four-fermion term in fact arises
in the effective action due to the interplay of higher- and low-momentum
lattice fermionic modes. This is discussed in section 7.  Before, we
comment shortly on the lattice analytics and continuum limit for the pure
2D Ising model. In what follows we assume ferromagnetic case with the
homogeneous distribution of bond coupling parameters over the lattice.

%
%
\section{ The pure case (lattice)}

{} From (\ref{qpp1}) specified to the homogeneous distribution of
bond couplings, we obtain the Gaussian integral for the partition
function of the regular $2D$ Ising model:
\bea
Q =\int \prod\limits_{mn}^{}d\bar{c}_{mn}dc_{mn}
\exp\sum\limits_{mn}^{}\,\big[\,c_{mn}\bar{c}_{mn}
-t_1\,c_{m-1n}\bar{c}_{mn}-\,t_2\,c_{mn}\bar{c}_{mn-1}
\nonumber \\
-\,t_1t_2\,c_{m-1n}\bar{c}_{mn-1} + t_1\,c_{mn}c_{m-1n} +
t_2\,\bar{c}_{mn-1}\bar{c}_{mn}\,\big]\,,
\label{pure1}
\eea
where $t_{1,2}=\tanh\,b_{\,1,2}\,$, $\,b_{\,1,2}= J_{1,2}/kT$. Let us pass
to the momentum space by Fourier substitution:
\bea
c_{mn}=\frac{1}{\sqrt{L^{\,2}}}\,\sum\limits_{pq}^{}
c_{pq}\,\eee^{-\,i\frac{2\pi p}{L}m+ i\frac{2\pi q}{L}n}\,,\;\;\;
\bar{c}_{mn}^{}\,=\,\frac{1}{\sqrt{L^{\,2}}}\,\sum\limits_{pq}
\bar{c}_{pq}^{}\,\eee^{\,+i\frac{2\pi p}{L}m-i\frac{2\pi q}{L}n}\,,
\label{fur1}
\eea
with $-L/2 \leq p,q \leq L/2$, where $L$ is lattice length.
The partition function becomes:
\bea
Q =\int \prod\limits_{pq}^{} d\bar{c}_{pq}^{}dc_{pq}^{}\,
\exp\,\sum\limits_{pq}^{}\,\Big[\,(1-t_1\,\eee^{\,i\frac{2\pi
p}{L}}-t_2\,\eee^{\,i\frac{2\pi q}{L}}-t_1t_2\,\eee^{\,i\frac{2\pi p}{L}
+ i\frac{2\pi q}{L}}) \,c_{pq}^{}\bar{c}_{pq}^{}
\nonumber \\
+\, t_1\,\eee^{\,i\frac{2\pi p}{L}}\,c_{L-pL-q}^{}c_{pq}^{}
+\,t_2\,\eee^{i\frac{2\pi q}{L}}\,\bar{c}_{pq}^{}
\bar{c}_{L-pL-q}^{}\,\Big]\,.
\label{qpp3}
\eea
Making the $pq \leftrightarrow L-pL-q$ symmetrization of the sum (needed
since variables with momenta $pq$ and $L-pL-q$ interact) we find the
integral (\ref{qpp3}) to be decoupled into a product of simple
low-dimensional integrals, each over $d\bar{c}_{pq}dc_{pq}d\bar{c}_{L-pL-q}
dc_{L-pL-q}$. The $pq$-integral factor can be readily evaluated making use
of the elementary rules, see (\ref{gr2}), which results in the explicit
solution for $Q^{\,2}$. For similar momentum-space calculations see also
\cite{ple88}. Taking then the limit of infinite lattice, for the
free energy per site we obtain:
\bea
-\beta f_Q =
\frac{1}{2}\int\limits_{\!-\pi}^{\pi}\!\int\limits_{\!-\pi}^{\pi}
\frac{dp\,dq}{(2\pi)^{\,2}}\ln \Big[(1+t_{1}^{2})(1+t_{2}^{2})
-2t_1(1-t_{2}^{2})\cos p-2t_2(1-t_{1}^{2})\cos q \Big]\,,
\label{free1}
\eea
which is equivalent to the Onsager solution \cite{ons44}.
As it follows from the analysis of the free-energy form, in the
ferromagnetic case, $t_1,t_2>0$, the critical point is given by the
condition:
\bea
1-t_1-t_2\,-\,t_1\,t_2\,=\,0\,.  \triex
\label{crit1}
\eea
The specific heat exhibits the logarithmic singularity near the critical
point, $C\simeq A_c|\ln|\tau||$, $\tau\sim|T-T_c\,|\to 0$, where $A_c$ is
the critical amplitude. For the symmetric lattice ($t_1=t_2$) the amplitude
is:  $A_c= (8/\pi)\,b_{c}^{\,2}$, where $b_{c}= \frac{1}{2}\ln(1+\sqrt{2})
=(J/kT_c)$ is the dimensionless inverse critical temperature.

%
%
\section{ Majorana and Dirac Fields}

Let us make a comment on the continuum-limit interpretation of the
pure 2D Ising model near the critical point, also see \cite{drouf89,ple95}.
It is interesting that already the exact lattice action from (\ref{pure1})
can be written in the field-theoretical like form, cf. \cite{ple95}. In
terms of the lattice derivatives $ \pal_m c_{mn} = c_{mn}-c_{m-1n},\,
\pal_n \bar{c}_{mn} =\bar{c}_{mn}-\bar{c}_{mn-1}$, this action becomes:
\bea
S\,(c,\bar{c}\,)_{\rm\,pure} =
\sum\limits_{mn}^{}\,\Big[\,\underline{m}\,
c_{mn}\bar{c}_{mn} +\lambda_1\,  (\pal_m c_{mn})\bar{c}_{mn}
+\,\lambda_2\,c_{mn}(\pal_n\bar{c}_{mn})
\nonumber \\
+\,t_1\,(\pal_m c_{mn})c_{mn}+\,t_2\,\bar{c}_{mn}(\pal_n \bar{c}_{mn})
-t_1t_2\,(\pal_m c_{mn})\,(\pal_n \bar{c}_{mn})\,\Big]\,,
\label{major1}
\eea
with $\underline{m}= 1- t_1 -t_2 -t_1\,t_2,\; \lambda_{1}^{} = t_1 +
t_1\,t_2\,,\; \lambda_{2}^{} =t_2+t_1\,t_2$. Evidently, $\underline{m}$
plays the role of mass, the critical point corresponds to
$\underline{m}=0$. Taking the formal limit to the continuum space: $mn \to
(x_1,x_2)= x,\; \pal_m, \pal_n \to \pal_1 ,\pal_2$, and changing the notion
for the fields:  $c_{mn}, \bar{c}_{mn} \to \psi(x), \bar{\psi}(x)\,$, we
obtain the continuum-limit counterpart of the exact lattice action
(\ref{major1}) of the same formal structure.  The second order kinetic term
with $\pal_1\pal_2$ can be ignored. The remaining action is of the Majorana
type and can be transformed into a canonical form by a suitable rotation
and rescaling of the fields, $\psi, \bar{\psi} \to \psi_1,\psi_2$, cf.
\cite{drouf89,ple95}.  In this way we obtain the action in a canonical
Majorana form:
\bea
S\,(\psi_1,\psi_2) = \int d^2x\,\big[\,\ovl{m}\,\psi_1\psi_2 +
\,\psi_1\mbox{$\frac{1}{2}$}(\pal_1+i\,\pal_2)\psi_1 +
\psi_2\mbox{$\frac{1}{2}$}(-\pal_1+i\,\pal_2)\psi_2\,\big]\,,
\label{major3}
\eea
with the rescaled mass:
\bea
\ovl{m}= \frac{1-t_1-t_2-t_1\,t_2}{\sqrt{2\,(t_1t_2)_c}}\,,
\label{mass2}
\eea
where $(\ )_c$ stands for the criticality condition, eq. (\ref{crit1}).

The 2D Majorana action (\ref{major3}) can be rewritten also in the matrix
notation. Assuming $\sigma_1,\sigma_2,\sigma_3$ to be the standard Pauli
matrices, we find:
\bea
S_{\rm\,major} = \mbox{$\frac{1}{2}$}\,\int
d^{\,2}x\;\tilde{\Psi}\,(x)\,[\;\ovl{m}\, +
\hat{\pal}\,]\;\Psi^{}\,(x)\;, \;\;\;
\tilde{\Psi}\,(x) = \Psi^{\,T}\,(x)\,(i\,\sigma_2)\,, \;\;\;
\hat{\pal}=\gamma_1\pal_1+\gamma_2\pal_2\,,
\label{major5}
\eea
where $\tilde{\Psi}$ and $\Psi$ are the two-component left and right
Majorana spinors, $\Psi =(\psi_1,\psi_2)$, and $\gamma_1=\sigma_1$,
$\gamma_2=\sigma_2$ are the 2D gamma-matrices. Finally, we can pass to the
Dirac theory of charged fermions by doubling the number of fermions in the
Majorana representation \cite{shal94,ple95}. Taking two identical copies
$S'$ and $S''$ of the Majorana action, eq. (\ref{major5}), we consider the
combined action $S_{\rm\,dirac} =(S'+S'')_{\rm\, major}\,$. Introducing
new Dirac fields by substitution $\Psi =\frac{1}{\sqrt{2}}\,
\big(\Psi^{\,'} +i\,\Psi^{\,''} \big),\, \bar{\Psi} = \frac{1}{\sqrt{2}}\,
\big(\tilde{\Psi}^{\,'} -i\,\tilde{\Psi}^{\,''}\big)\,$, we find the
combined action in the form:
\bea
S_{\rm\,dirac} = \,\int d^{\,2}x\;\bar{\Psi}\,(x)\,
[\,\ovl{m}\, +\hat{\pal}\,]\;\Psi^{}\,(x)\;,
\label{dirac1}
\eea
where $\ovl{m}$ and $\hat{\pal}$ are the same as in (\ref{major5}),
while  $\bar{\Psi}$ and $\Psi$ are now charged Dirac spinors with four
independent components.

The formal meaning of the transformation to the continuum limit can be
most readily understood in the momentum space. The change $\pal_m,\pal_n
\to  \pal_1,\pal_2\,$ in the momentum space corresponds to the
approximation:
\bea
\ba{llr} \eee^{\,ip}-1\simeq\; ip\,, \;\;\;
\eee^{\,iq}-1\simeq\; iq\,,
\ea\label{exp1}
\eea
where $p,q$ are the quasidiscrete Fourier momenta, $p,q \leftrightarrow
\frac{2\pi p}{L},\frac{2\pi q}{L}$, $-\pi \leq p,q \leq \pi$, see
(\ref{fur1}), and $|p|$ below is modulus of $(p,q)$. The approximation
like (\ref{exp1}) is legal only in the neighborhood of the origin in the
momentum space, $0\leq |p|\leq k_0$, where $k_0 \ll 1$ is some fixed small
cut-off momentum. There is also an effective infrared bound for $|p|$
introduced by the mass, which implies that $\ovl{m}$ is small and we are
near $T_c$. The continuum-limit formulation thus captures the essential
features of the exact lattice theory in the low-momentum sector responsible
for the singularities in the thermodynamic functions and the large-distance
behaviour of the correlations as $T\to T_c$ ($\ovl{m} \to 0$).
In this region, the approximation is exact. In particular, the logarithmic
singularity in the specific heat can still be recovered from (\ref{major5})
or (\ref{dirac1}), with the exact value of the critical amplitude.

%
%
\section{ Gross-Neveu model ($N \to 0$)}

To extract the continuum-limit theory for the model with interaction
(\ref{qnn50})-(\ref{sint1}), let us distinguish explicitly the lattice
modes, cf.~(\ref{fur1}), with higher and low fermionic momenta,
$|p|>k_0$ and $|p|<k_0$, where $k_0 \ll 1$ is small fixed cut-off.
Introducing separable notation $a,\psi$ for the corespondent fields,
in the coordinate space we write:
\bea
c_{mn}= (a_{mn})_{\;|\,p\,|>k_0} + (\psi_{mn})_{\;|\,p\,|<k_0}\,,
\;\;\;
\bar{c}_{mn}= (\bar{a}_{mn}) _{\;|\,p\,|>k_0} +
(\bar{\psi}_{mn})_{\;|\,p\,|<k_0}\,.
\label{decomp1}
\eea
The Gaussian part of the exact lattice action (\ref{qnn80}) is additive in
the momentum space, so $S_0=S_a +S_\psi$, where $S_a$ and $S_\psi$ are
independent.  However, the  non-Gaussian term of interaction (\ref{sint1})
is non-additive with respect to $c \to a +\psi$. We have thus to substitute
explicitly $c \to a + \psi$ into this term, and then integrate over the
higher-momentum lattice modes $(a_{mn}) _{\,|p|>k_0}$. This is done below
for weak dilution, in the first order of perturbation in $\lambda_0 \sim
1-p \ll 1$.

Writing the action (\ref{qnn80}) in the form $S=S_{\,a}+ S_{\,\psi}
+ \lambda_0\,S_{\,int}\,(a,\psi)$, $\lambda_0=\frac{1-p}{p}$, in the
first order in $\lambda_{0}$ we obtain:
\bea
&& \overline{\stackrel{}{Q^{\,N}}}
\simeq \int D_{\,a}\,D_{\,\psi}\;\eee^{\,S_a+S_\psi}\,
\big( 1 + \lambda_0\,S_{\,int}\,(a,\psi)\big)
\nonumber \\
&& \simeq \int D_{\,a}\;\eee^{\,S\,(a)}\int D_{\,\psi}\,\eee^{\,S_\psi}\,
\left\{ \frac{ \int D_{\,a}\eee^{\,S_{\,a}}
\big(1+ \lambda_0\,S_{\,int}\,(a,\psi)\big)}
{  \int D_{\,a}\eee^{\,S_{\,a}}}\right\}
\nonumber \\
&& \simeq \int{D}_{\,a}\;\eee^{\,S_{\,a}} \int D_{\,\psi}\,\eee^{\,
S_\psi + \lambda_{\,0}\, \left<S_{\,int}\,(a,\psi)\right>_{a}\,}\;.
\label{rrr2}
\eea
Substituting (\ref{decomp1}) into $\left<S_{\,int}(a,\psi)\right>_a$, we
follow the standard field-theoretical prescriptions for critical
fluctuations that only the most important quartic and quadratic terms in
the $\psi, \bar{\psi}$ fields are actually to be taken into account
\cite{shal94,vikd95}. In terms of the characteristic fermionic averages
$\left<A\right>$ and $\left<B\right>$ presenting the contribution of the
higher-momenta lattice modes, we find:
\bea
&& \left<\,S_{\,int}\,(a,\psi)\right>_{a}=
\,\sum\limits_{mn}^{}\prod\limits_{\alpha=1}^{N} \Big[\,\left<A\right>
+\left<B\right>\,\psi_{mn}^{\alpha}\bar{\psi}_{mn}^{\alpha}+ ...\, \Big]\,
\nonumber \\
&&= \left<A\right>^{N}\, \sum\limits_{mn}^{} \bigg[\, 1 +
\frac{\left<B\right>} {\left<A\right>}\,\sum\limits_{\alpha=1}^{N}
\,\psi_{mn}^{\alpha}\bar{\psi}_{mn}^{\alpha}+
\frac{1}{2}\,\frac{\left<B\right>^2}{\left<A\right>^2}\,
\Big(\sum\limits_{\alpha=1}^{N}
\psi_{mn}^{\alpha}\bar{\psi}_{mn}^{\alpha}\Big)^{\,2}
+ ...\,\bigg]\,,
\label{rrr7}
\eea
where
\bea
&& \left<A\right> =
\left< a\bar{a}\,\exp\,(t_1t_2\,\pal_1a\,\pal_2\bar{a})\right>_{a}
=\left< a\bar{a}\right>_{a} +
t_1t_2\,\left<a\bar{a}\,\pal_1a\,\pal_2\bar{a})\right>_{a}\,,
\nonumber \\
&& \left<B\right>
= \left< \exp\,(t_1t_2\,\pal_1a\,\pal_2\bar{a})\right>_{a}
= 1 + t_1t_2\,\left<\pal_1a\,\pal_2\bar{a})\right>_{a}\,.
\label{aveab1}
\eea
These are the lattice averages for pure model (\ref{pure1}) with infrared
truncation at $k_0$. Here we assume the evident abbreviation with dropped
$_{mn}^{\alpha}$ index and $\pal_1, \pal_2 \leftrightarrow \pal_m,\pal_n$
correspondence. For instance, $\left<\pal_1a \pal_2\bar{a}\right>_a$ means
$\left<\pal_m a_{mn}^{\,\alpha} \pal_n\bar{a}_{mn}^{\,\alpha}\right>_a\,$,
etc.  The dependence on small momentum cut-off $k_0$ can be in fact ignored
(put $k_0=0$). The averages are then to be evaluated with the pure lattice
action, eqs.  (\ref{pure1})-(\ref{qpp3}), and are $_{mn}^{\alpha}$
independent.

In (\ref{rrr7}), in the first line  $(...)$ is the abbreviation for
the `gaussian kinetic terms', which can be ignored, and in the second line
$(...)$ stands for the total sum of all irrelevant corrections, including
the gaussian kinetic terms, quartic kinetic terms, and higher-order
corrections in the $\psi,\bar{\psi}$ fields of any kind.  The effective
action for the $\psi,\bar{\psi}$ fields is $S_{\rm eff} = (S_{\psi})_{\rm
free} + \lambda_0 \left<S_{\rm\,int }^{}(a,\psi)\right>_{a}$. The terms
preserved explicitly in (\ref{rrr7}) contribute to the Gaussian part of
this action (modification of mass) and generate the four-fermion
interaction. A trivial constant term may be ignored. The Gaussian part of
the action $S_{\rm eff}$ appears in the form of the replicated
continuum-limit version of the action (\ref{major1}), cf. (\ref{major3}),
but with a modified mass. After uniformization of the Gaussian part and the
correspondent rescaling of the fields, $\psi,\bar{\psi} \to \psi_1,\psi_2$,
we obtain the effective continuum-limit action in the form of the
$N$-colored Gross-Neveu model \cite{dd83,shal94}:
\bea
S_{\rm\,G-N} =\int d^2x \Big\{\sum\limits_{\alpha=1}^{N}\Big[\,
\ovl{m}_N\,\psi_{1}^{\alpha}\psi_{2}^{\alpha}
+\,\frac{1}{2}\,\psi_{1}^{\alpha}\,(\pal_1+i\pal_2)\,\psi_{1}^{\alpha}
\nonumber \\
+ \frac{1}{2}\,\psi_{2}^{\alpha}\,(-\pal_1+i\pal_2)\,
\psi_{2}^{\alpha}\Big] + g_{\mathstrut N} \Big[\sum\limits_{\alpha=1}^{N}
\psi_{1}^{\alpha}\psi_{2}^{\alpha}\Big]^{\,2}\,\Big\}\,,
\label{gross1}
\eea
with the effective mass and coupling constant given as follows:
\bea
&& \ovl{m}_{\mathstrut N}\,
=\,\frac{1-t_1-t_2-t_1\,t_2}{\sqrt{2(t_1t_2)_c}} +
\left<A\right>^N\,
\frac{1-p}{p}\,\frac{\left<B\right>}{\left<A\right>}\,
\frac{1}{\sqrt{2\,(t_1t_2)_c}}\,,
\label{mnn1}
\\
&& g_{\mathstrut N}\, = \left<A\right>^N\,
\frac{1-p}{p}\,\frac{\left<B\right>^2}{\left<A\right>^2}\,
\frac{1}{4\,(t_1t_2)_c}\,.
\label{gnn1}
\eea
Since the replica limit $N \to 0$ is assumed at final stages, one can put
$N=0$ and $\left<A\right>^{N}=1$ already in (\ref{mnn1}) and (\ref{gnn1}).
In the matrix notation, the action (\ref{gross1}) becomes:
\bea
S_{\rm\,G-N} =
\frac{1}{2}\,\sum\limits_{\alpha=1}^{N}\,\int d^2x\,
\tilde{\Psi}^{\alpha}(x)\,(\,\ovl{m}_{\mathstrut 0}+ \hat{\pal}\,)\,
\Psi^{\alpha}(x)+
g_{\mathstrut 0}\,\int d^2x\,\Big[ \frac{1}{2}\sum\limits_{\alpha=1}^{N}
\tilde{\Psi}^{\alpha}(x)\Psi^{\alpha}(x)\,\Big]^{\,2}\,.
\label{gross2}
\eea
The Gaussian part here is the replicated Majorana action (\ref{major5})
with modified mass (\ref{mnn1}). By analogy with the pure case, we can
pass as well to the Dirac field interpretation. The doubling
of Majorana fermions can be realized simply by doubling the number of
replicas, $N \to 2N$, and we find:
\bea
S_{\rm\,G-N} = \sum\limits_{\alpha=1}^{N}\,\int d^2x\,
\bar{\Psi}^{\alpha}(x)\,(\,\ovl{m}_{\mathstrut 0} + \hat{\pal}\,)\,
\Psi^{\alpha}(x)+
g_{\mathstrut 0} \,\int d^2x\,\Big[ \sum\limits_{\alpha=1}^{N}
\bar{\Psi}^{\alpha}(x)\Psi^{\alpha}(x)\,\Big]^{\,2}\,,
\label{gross3}
\eea
where $\bar{\Psi}^{\,\alpha}, \Psi^{\,\alpha}$ are charged Dirac
spinors, cf.  (\ref{dirac1}).  In (\ref{gross2}), (\ref{gross3}) we
assume the mass and charge already at $N=0$.

The $N=0$ Gross-Neveu model was first introduced as an effective field
theory for weak {\em bond}\, dilution in the pioneering papers by Dotsenko
and Dotsenko (DD) \cite{dd81}-\cite{dd83}. Their theory has been reanalyzed
later on as regards to the behaviour of correlations by Shalaev, Shankar,
and Ludwig (SSL) \cite{shal84}-\cite{lud90}, see \cite{shal94} for a
comprehensive review. An advanced analysis of correlations in the $N=0$
G-N model in terms of the perturbed conformal field theories has been
performed quite recently by Dotsenko, Picco and Pujol \cite{dpp95,dpp96}.
We thus expect the same changes in critical behaviour inspired by weak {\em
site} dilution as those predicted in the DD-SSL theories for weak bond
dilution.  From the renormalization-group (RG) analysis of the $N=0$ G-N
model it follows that in a closed vicinity of the critical point the
singularity in the specific heat is doubled-logarithmic \cite{dd83,vikd95}:
\bea
C \propto  \frac{1}{2g_{0}^{}} \ln\Big(1+
\frac{2g_{0}^{}}{\pi}\ln\frac{1}{|\tau|}\Big)\,.
\label{heat5}
\eea
The RG calculation also shows \cite{dd83} that the  correlation radius
gains the logarithmic correction, $R_c \sim \tau^{-1}|\ln \tau|^{1/2}$ (cf.
$R_c \sim \tau^{-1}$ in the pure case, where $\tau \sim |T-T_c|$). The
two-spin correlation function $G\,(R) = \left<\sigma(0)\sigma(R)\right>$
taken exactly at $T_c$ remains to be the same as for the pure system
\cite{shal94}: $G\,(R) \sim R^{\,-1/4}$, $R \to \infty$. [That no
logarithmic or other corrections appear due to the disorder in $G(R)$ in
the $N=0$ G-N model can now be assumed to be definitely established
\cite{shal94}, \cite{dpp95}, \cite{dpp96}, this question was a subject of
some debate in previous decade \cite{shal84}-\cite{lud90}]. Assuming the
standard scaling arguments to be true, this then implies that the
magnetization and magnetic susceptibility gain the logarithmic corrections
to the pure-case power singularities at criticality \cite{shal94,SST95}:
$M\,(\tau\,|\,0) \sim |\tau|^{1/8}|\ln \tau|^{-1/16},\; \chi\,(\tau\,|\,0)
\sim |\tau|^{-7/4}|\ln \tau|^{7/8},\; M\,(0\,|\,h) \sim h^{1/15}$, where
$\tau \sim |T-T_c|$ and $h$ is a small magnetic field. These predictions
have been checked also, for weak dilution, in a series of the precise
Monte-Carlo experiments \cite{wsdoa90p}, \cite{adosw90n}, \cite{schtal94},
\cite{SST95}. The question about the effects produced by strong and
moderated dilution is less evident, however, both on theoretical
\cite{zieg90,shal94} and experimental (Monte-Carlo) \cite{kim94,SST95}
sides, this holds for both bond and site dilution.

In conclusion to this section, let us specify in more detail the parameters
of the $N=0$ G-N model for site dilution for the isotropic lattice
($t_1=t_2$, or $J_1= J_2$). This will yield also the exact value for the
initial slope of the $T_c(p)$ curve.  The effective mass and coupling
constant include the lattice constants $\left<A\right>$ and
$\left<B\right>$. Since the dilution is weak, and we are near $T_c$ of the
pure system, it is reasonable to specify these parameters exactly at $T_c$
of the pure system.  Also, the dependence on small cut-off momenta, $k_0
\ll 1$, can be ignored. The averages $\left<A\right>$ and $\left<B\right>$
are then to be taken for the pure lattice system at $T_c$. If $t_1\neq
t_2$, however, these parameters still remain be functions of lattice
anisotropy. They become definite numbers in the isotropic case. By a
straightforward though somewhat lengthy calculation we have obtained the
following values for the isotropic lattice ($J_1=J_2,\,T=T_c,\,k_0=0$):
\bea
&& \left<A\right>_{c} =
\frac{1}{2}\,\bigg(\frac{1}{\sqrt{2}}+\frac{1}{\pi}\bigg)
\bigg(1+\frac{1}{\sqrt{2}} -\frac{1}{\pi}\bigg) =0.\,712048\,,
\nonumber \\
&& \left<B\right>_{c}= 2\,t_c\,\bigg(1+\frac{1}{\sqrt{2}}-
\frac{1}{\pi}\bigg) =1.\,150517\,, \;\;\; t_c=\sqrt{2}-1\,,
\nonumber \\
&& \frac{\left<B\right>_{c}}{\left<A\right>_{c}}=
4\,\sqrt{2}\,t_c\,\bigg(1 + \frac{\sqrt{2}}{\pi}\bigg)^{-1}
=1.\,615786\,.
\label{aveab2}
\eea
For the isotropic lattice, the critical point is given by condition
$(1-2t-t^2)_c=0$, with the solution $t_c= \eee^{-2b_c}= \sqrt{2}-1$,
$b_c=\frac{1}{2} \ln(1+\sqrt{2})$. The pure-case mass $\ovl{m}=
\ovl{m}(1)$ (\ref{mass2}) then appears in the form: $\ovl{m}\,(1)
\simeq 2(t_c-t)/t_c \simeq 4b_c(T-T_c)/T_c$. Taking into account
(\ref{aveab2}), the effective mass ($N=0$) of the G-N model becomes:
\bea
\ovl{m}_{\mathstrut 0} = 4\,b_c\,\frac{T-T_c}{T_c}
+\frac{1-p}{p}\,\frac{4}{1+\sqrt{2}/\pi}
\label{mass7}
\\ \nonumber
=(1.763)\,\frac{T-T_c}{T_c}+(2.758)\,\frac{1-p}{p}\,,
\eea
while the coupling constant ($N=0$) appears in the form:
\bea
g_{\mathstrut 0}=\frac{1-p}{p}\,\frac{8}{(1+\sqrt{2}/\pi)^{\,2}}
\simeq (3.804)\,\frac{1-p}{p}\,.
\label{goo1}
\eea
These are the values of mass and charge of the $N=0$ Gross-Neveu model
(\ref{gross1})-(\ref{gross3}) for weak site dilution near $T_c$ in
the isotropic case. These values are exact in the linear orders in
$1-p$ and $T-T_c$.

The expression for mass (\ref{mass7}) enables one to define the critical
slope for the $T_c(p)-p$ curve at $p=1$. The initial slope is the
coefficient in the expansion $T_{c}^{}(p) = T_{c}^{}(1)\,[\,1-S_c\,
\frac{1-p}{p} + \ldots\,]\,$.  Dividing both sides of the equation
$\ovl{m}_{0}(p)=0$ by $1-p$ and taking the limit $p\to 1$, we find:
\bea
S_c = \left.\frac{1}{T_c}\,\frac{d\,T_c}{d\,p}\,\right|_{\,p=1}=
\frac{(b_c)^{-1}}{1+\sqrt{2}/\pi}=1.564785\,,
\label{slope1}
\eea
which is in agreement with the earlier calculation of $S_c$ \cite{auyang76}.
As distinct from a direct spin-lattice perturbative analysis in
\cite{auyang76}, the present calculation includes the $N$-replica trick with
fermions.

%
%
\section{Concluding discussion}

We have applied a new noncombinatorial method of fermionization to clarify
the fermionic structure of the 2D Ising model with quenched site dilution.
In fact, the method works equally well for both site and bond dilution. We
have specified the discussion merely to the case of site dilution as being
less studied theoretically. For the first step, the partition function of
the 2DIM with fixed site dilution and arbitrary inhomogeneous distribution
of the bond coupling parameters over the lattice was transformed into a
fermionic Gaussian integral, eq. (\ref{qab1}). The resulting integral was
then simplified integrating out extra fermionic degrees of freedom. This
yields the Gaussian like representation with two fermionic variables per
site for fixed site dilution, eqs. (\ref{qcc1})-(\ref{qpp1}). Even for the
pure case, this reduction of extra fermionic degrees of freedom essentially
simplifies the analytics of any kind and also illuminates Majorana-Dirac
structures in the pure 2DIM already at lattice level. The quenched
averaging over the disorder in the $N$-replica scheme results, in
turn, in the exact lattice theory of interacting fermions for the case
of site dilution, eqs.  (\ref{qnn50})-(\ref{sint1}). The interaction in
this  exact lattice theory appears to be of order $2N$ in fermions, where
$N$ is the number of replicas. However, the continuum-limit approximation
for weak dilution near criticality results in the standard $N=0$
Gross-Neveu model with four-fermion interaction. The $N=0$ G-N model has
been already intensively analyzed as an effective field theory for bond
dilution \cite{dd83, shal94}.  This then implies similar behaviour of the
thermodynamic functions and correlations in the disordered $2D$ Ising model
both for site and bond dilution near $T_c$, at least for small amount of
impurities. The predictions of the $N=0$ G-N model for weak dilution are
the log-log singularity in the specific heat and log-corrections in $M$ and
$\chi\,$, see section 7 for more comments. We have  also specified in
detail the parameters (mass and charge) of the $N=0$ G-N model for site
dilution in terms of the lattice fermion Green's functions and evaluated
the initial slope of the $T_c-p$ curve.  The present analysis thus supports
the hypothesis about the universality in critical behaviour produced by
different sorts of disorder for a small fraction of impurities in the
ferromagnetic 2D Ising model. The question about the universality between
weak and strong dilution remains to be open, in essence. A more
straightforward analysis, directly in lattice interpretation, seems to be
needed. A compact form of interaction in the exact lattice theory
(\ref{qnn50})-(\ref{sint1}) provides grounds to suppose further progress in
this respect. Finally, let us note that the Gaussian fermionic
representations with fixed disorder like (\ref{qab1}), (\ref{qcc1}),
(\ref{qpp1}) yet preserve all the information about the original
spin-variable model (\ref{qss1}) that is included in the hamiltonian
(\ref{ham1}). There are no also any restrictions, at this stage, on the
sign of the bond coupling parameters. Therefore, these representations can
be used as the starting point to try other problems with disorder in the
$2D$ Ising model, beyond site or bond dilution (spin glasses, models
with annealed or regular inhomogeneities, etc). As well, one can try to
apply other methods of quenched averaging, beyond the $N$-replica scheme.
The later possibility may be favorable concerning the problem of strong
and moderated dilution.

\section*{Acknowledgments:}

I am especially indebted to Boris N. Shalaev for valuable comments at
the first stages of this work. The discussions with R. Hayn, G. Jug, W.
Selke, B. N.  Shalaev, A. I. Sokolov and K. Ziegler are gratefully
acknowledged.

\end{document}